\documentclass[twocolumn, twocolappendix, float]{aastex631} %linenumbers
\usepackage{graphicx} % Required for inserting images
\usepackage{amsmath}
\usepackage{mathtools}
\usepackage{enumitem}
\usepackage{soul}
\usepackage[normalem]{ulem}
\usepackage{CJK}
\usepackage{mathdesign}
\usepackage{hyperref}
\hypersetup{
    colorlinks=true,
    linkcolor=blue,
    filecolor=magenta,      
    urlcolor=blue,
    pdftitle={What does the stars say?},
    pdfcreator = {},
    pdfproducer = {}
    }

\newcommand{\axvr}{\textsf{arXiverse}}
\newcommand{\axv}{\textsf{arXiv}}
\newcommand{\aph}{\texttt{astro-ph}}
\newcommand{\aga}{\texttt{astro-ph.GA}}
\newcommand{\asr}{\texttt{astro-ph.SR}}
\newcommand{\aco}{\texttt{astro-ph.CO}}
\newcommand{\aep}{\texttt{astro-ph.EP}}
\newcommand{\ahe}{\texttt{astro-ph.HE}}
\newcommand{\aim}{\texttt{astro-ph.IM}}
\graphicspath{{./}{figures/}}
\shorttitle{Freeze or Crunch?}
\shortauthors{Tan}

\begin{document}
\begin{CJK*}{UTF8}{gbsn}

\title{Written in the Stars: How your (pens and) papers decide the fate of the \textsf{arXiverse}}

%\title{Let's talk about cheeky titles: I am NOT digging it (\textit{no, but actually yes})}
%\title{As a matter of colon: I am NOT digging cheeky titles (\textit{no, but actually yes $:>$})}
%\correspondingauthor{Joanne Tan}
%\email{joanne.tan@berkeley.edu}

%\author{Tie Sien Suk}
%\affiliation{34.413977, -119.842951}

\author{Joanne Tan}
\altaffiliation{The author is responsible for the contents of this paper, which do not in any way represent the views of their employers. Their views are their own, no matter if scrutinized through a microscope or a telescope.}
%\affiliation{You See Broccoli Memes for Edgy Teens.}
\affiliation{Earth, Maybe. Perhaps an Extragalactic Existence, or the Equivalent of a Hocus-pocus (EMPEE-EH).}
\altaffiliation{Read the acronym aloud to guesstimate where the author is based.}

\begin{abstract}
We all \textit{love} the ecstasy that comes with submitting papers to journals or \textsf{arXiv}. Some have described it as yeeting their back-breaking products of labor into the void, wishing they could never deal with them ever again. The very act of yeeting papers onto \textsf{arXiv} contributes to the expansion of the \textsf{arXiverse}; however, we have yet to quantify our contribution to the cause. In this work, I investigate the expansion of the \textsf{arXiverse} using the \textsf{arXiv} \texttt{astro-ph} submission data from 1992 to date. I coin the term ``the \axvr \ constant", $a_0$, to quantify the rate of expansion of the \textsf{arXiverse}. I find that \aph\ as a whole has a positive $a_0$, but this does not always hold true for the six subcategories of \aph. I then investigate the temporal changes in $a_0$ for the \aph\ subcategories and \aph\ as a whole, from which I infer the fate of the \axvr. %Using \aph\ as a whole as a proxy for the \axvr, the current $a_0$ is measured to be 0.110\pm0.002 paper/day/month. 
%From the temporal changes in $a_0$, 
I conclude that our \axvr\ is past its peak of expansion and could gradually slow down to a crunch.  
%What's in a name, a poet once asked. To which we present this work, where we investigate the importance of a paper title in ensuring its best outcome. We queried astronomy papers using NASA ADS and ranked 6000 of them in terms of cheekiness level. We investigate the correlation between citation counts and (i) the presence of a colon, and (ii) cheekiness ranking. We conclude that colon matters in the anatomy of a paper title. So does trying to be cheeky, but we find that too much cheekiness can lead to cringefests. Striking the right balance is therefore crucial. May we recommend aiming for a level 4 cheekiness on a scale of $1-5$. 
\end{abstract}

\section{Introduction}
\label{sec1:intro}
\axv\ is an open-access e-print archive that hosts multi-million scholarly articles in various fields. Researchers regularly submit their articles to \axv\ as a way to share their research with people around the world. \axv\ host articles from several main fields, namely physics, mathematics, computer science, quantitative biology, quantitative finance, statistics, electrical engineering and systems science, and economics. Each of these scientific fields also consists of multiple subfields. \axv\ was founded in August 1991 and has been growing steadily ever since. Honestly, if you are reading this paper, I don't think I need to explain \axv\ in more detail --- you are already on \axv!  

While \axv\ has been growing in popularity among researchers, there has not been an effort to quantify their contribution to the growth (or expansion) of \axv. I find this intriguing since we researchers \textbf{LOVE} to submit our papers into the void of \axv\ (and of journals, of course), but we never really stop to think about the consequences of our seemingly mundane and harmless action. We have derived so much euphoria from yeeting our paper to \axv, it is only right that we also consider the effects of our action. Therefore, in this paper, I aim to quantify our contribution to the expansion of the \axv. With that aim in mind, I set off to hunt down \axv's submission statistics, but only from the \aph\ category. This choice is motivated by
%I choose to use submission data from only the \aph\ category for 
two reasons: (i) it is the main category I yeet my research to; (ii) it was set up shortly after the birth of \axv\, and hence the data spans a temporal baseline of $\simeq$ 32 years. In \S\ref{sec2}, I describe the methods I have utilized to obtain the dataset necessary for this work, and the linear regression fits I perform on the datasets. In \S\ref{sec3:a0 const}, I discuss the rate of expansion of the \axv\ inferred from the model fits and the fact that it is not constant in time. Then, in \S\ref{sec4:disc}, I show robust proof that the rate of expansion changes over time, and discuss the limitations of this work, before delving into the future prospects of this work. I conclude the findings in \S\ref{sec6:conc}.

\section{Methods}
\label{sec2}
\subsection{Yoinking the required data}
\label{subsec2-1:yoink data}
I\footnote{Using the singular I in this paper because using we sounded really off in my head while writing.} utilize two query methods to obtain the data I need for this work. Firstly, I query the \axv\ via its API\footnote{\href{https://info.arxiv.org/help/api/index.html}{https://info.arxiv.org/help/api/index.html}} for papers submitted to \axv\ \aph\ from 1992 to 2008. I heavily modified the Python package \texttt{arxivscraper}\footnote{\href{https://github.com/Mahdisadjadi/arxivscraper}{https://github.com/Mahdisadjadi/arxivscraper}}\citep{arxivscrapper} to query for submitted papers using the \axv\ API instead of the \axv\ OAI-PMH interface\footnote{\href{https://info.arxiv.org/help/oa/index.html}{https://info.arxiv.org/help/oa/index.html}}. 
I filter for submissions to the \aph\ category from the queried metadata. Then, I select only the submissions with \aph\ as their primary category --- papers submitted to \aph\ as a secondary category are all excluded from the dataset. For every month from January 1992 to December 2008, I compute the number of papers submitted to \aph\ as their primary category and the number of unique submission dates. With these, I calculate the average daily submission rate for any given month. 

The second query method was using \texttt{arxivscraper} directly to query \axv\ for submission data from 2009 onwards. This was mainly because the first method was not able to handle the queries adequately, resulting in a time-consuming process to query the required metadata. The temporal split of the dataset in 2009 is due to the fact that \aph\ was growing exponentially and \axv\ decided to introduce six subcategories for \aph\, namely \aga, \asr, \aco, \aep, \ahe, and \aim, in January 2009. \texttt{arxivscraper} adequately handles queries efficiently for data since 2009+, for separate subcategories. Here, I apply a similar series of steps as the first method. I filter for submissions to the selected \aph\ subcategory, and then select only those that have the selected subcategory as their primary category. Then, for every month from January 2009 to February 2024, I calculate the number of papers submitted to the particular \aph\ subcategory and the number of unique submission dates. I then compute the average daily submission rate for the subcategory. To get the submission statistics for \aph\ as a whole, I sum up the number of papers submitted across the six subcategories to get the total submitted papers for the month. On the other hand, I use the number of days in the selected month as the number of unique submission dates. This decision is motivated by the fact that it is statistically very likely that at least one paper is submitted to \aph\ on any given day.\footnote{Yes, this is very handwavy, but 20-31 days are on the same order of magnitude anyway.} 
To ensure the robustness of this method, I compare the mined data to \axv's very own submissions statistics\footnote{\href{https://info.arxiv.org/about/reports/submission_category_by_year.html}{Arxiv's Submissions by Category since 2009+}} (available for submissions from 2009 onwards) and also a mini query using the first method for \aga. The comparisons show the data from the three sources/queries are highly similar. Therefore, despite the two different methods to query for submission statistics, I deduce that the two datasets can be combined temporally. From the queries, I obtain submission statistics for the following eight data groups:
\begin{enumerate} \itemsep -4pt
    \item \aga\ since 2009+
    \item \asr\ since 2009+
    \item \aco\ since 2009+
    \item \aep\ since 2009+
    \item \ahe\ since 2009+
    \item \aim\ since 2009+
    \item \aph\ since 2009+ (combination of groups 1--6)
    \item \aph\ since 1992+ (data from first query method, combined with group 7)
\end{enumerate}

\subsection{Fitting models to the data}
\label{subsec2-2:LR fits}
I perform a linear regression (LR) fitting onto the average daily submission rate as a function of time, by using \texttt{sklearn.linear\_model.LinearRegression} and \texttt{scipy.stats.linregress}. I first fit the LR model onto the complete dataset of a group (as defined at the end of Section \ref{subsec2-1:yoink data}) to obtain the slope and intercept of the LR line. Hereafter, I refer to this fit as LR-full. 
Then, I split the dataset into a training dataset and a testing dataset using the 80:20 split. I fit an LR model onto the training dataset and then use the model to generate predictions using the testing dataset. % --- this essentially means that I fit the LR model to 4/5 of the dataset but apply the model onto the full dataset. 
This fit is referred to as LR-split hereafter.
Next, I compare the slopes and intercepts of the LR fits on the full dataset and split dataset obtained from both \texttt{sklearn.linear\_model.LinearRegression} and \texttt{scipy.stats.linregress}; the results are found to be identical up to at least 6 s.f. Hence, I show only the results from \texttt{scipy.stats.linregress} as this function also provides the standard errors of the slopes and intercepts. Last but not least, I determine the 95\% confidence interval of the LR fits. At every time point, I draw 1000 random samples from a Gaussian distribution, with the slope as the center and the slope's standard error as the width. I repeat the same random draws for the intercept and its standard error. Every draw provides a predicted submission rate, giving me a distribution of 1000 predictions at every point in time. I then compute the 2.5th and 97.5th percentile of the distribution to form the 95\% confidence interval of the fit. 

\begin{figure}[h!]
\centering
\includegraphics[width=\columnwidth]{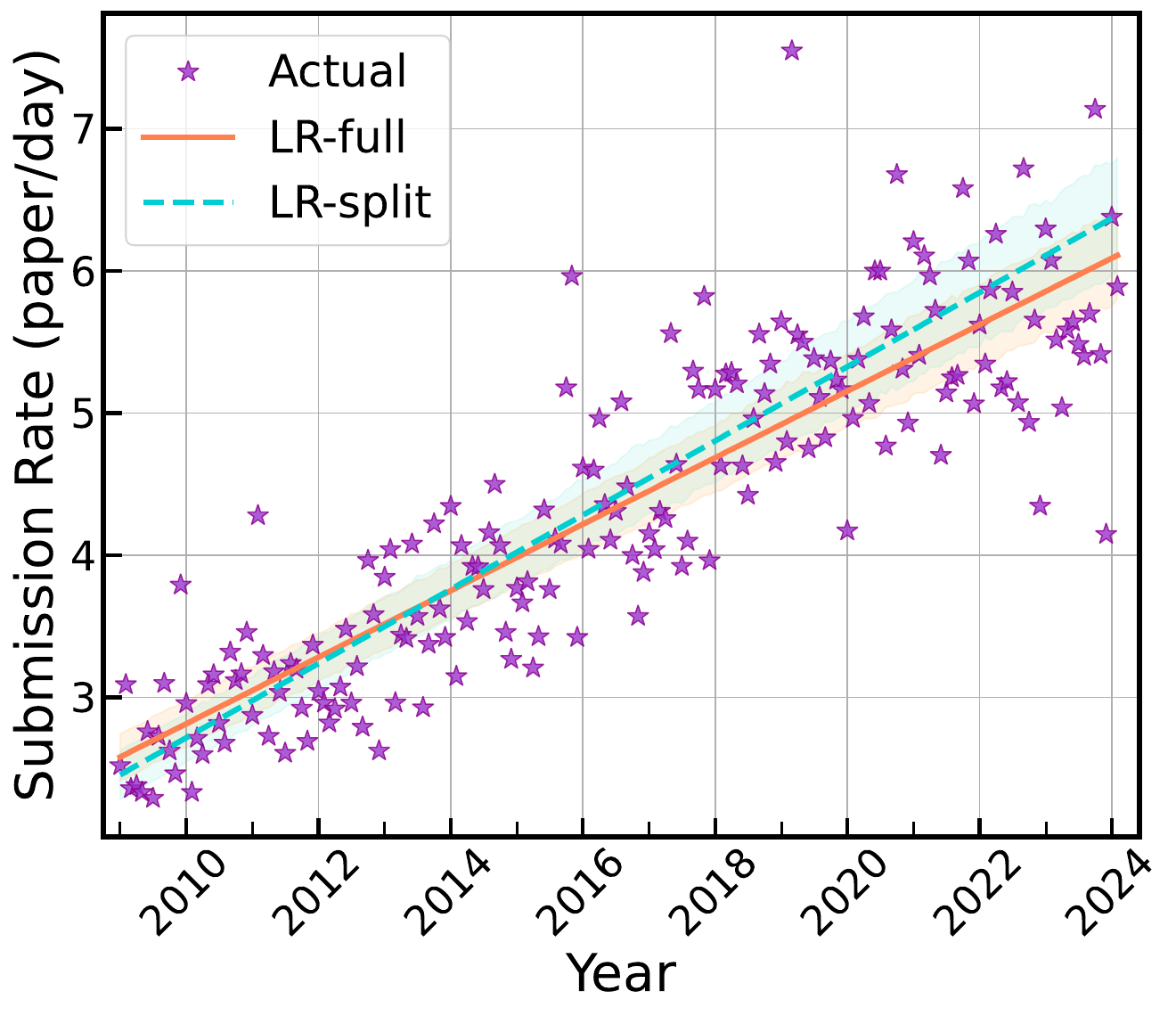}
\includegraphics[width=\columnwidth]{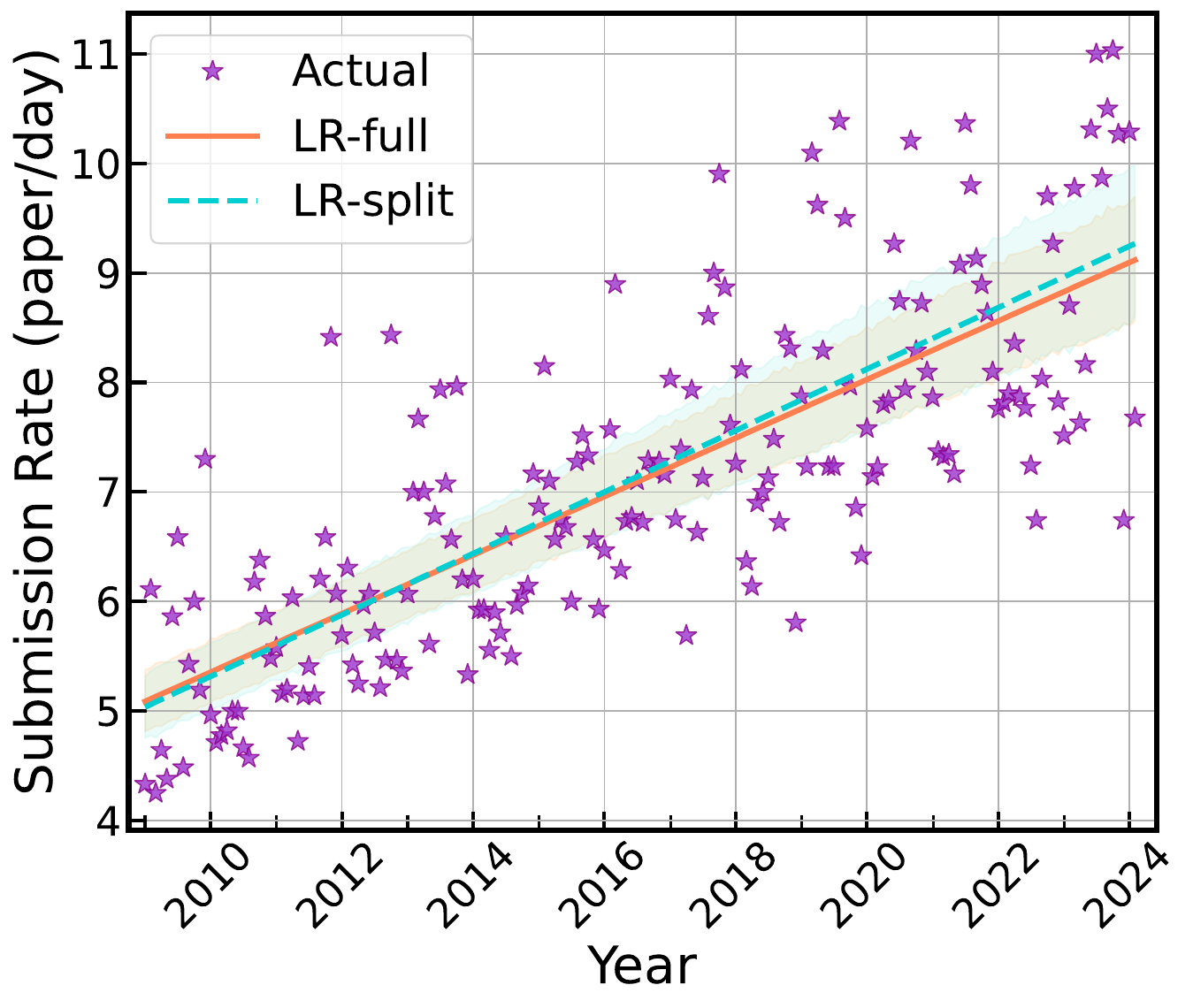}
\caption{\label{fig1:LR_fits_EP_HE} Submission rate as a function of time for \aep\ (group 4, top panel) and \ahe\ (group 5, bottom panel). In both panels, the purple stars indicate the actual submission rate in a given month. The orange line indicates the LR fit from the full dataset (LR-full), while the cyan dashed line indicates the LR fit from the split dataset (LR-split). The shaded regions around the LR fits indicate the 95\% confidence interval of the LR fit. The difference in the slopes of the LR-full and LR-split fits indicates a change in the expansion rate of the \axvr.}
\end{figure}

\begin{figure}[h!]
\centering
\includegraphics[width=\columnwidth]{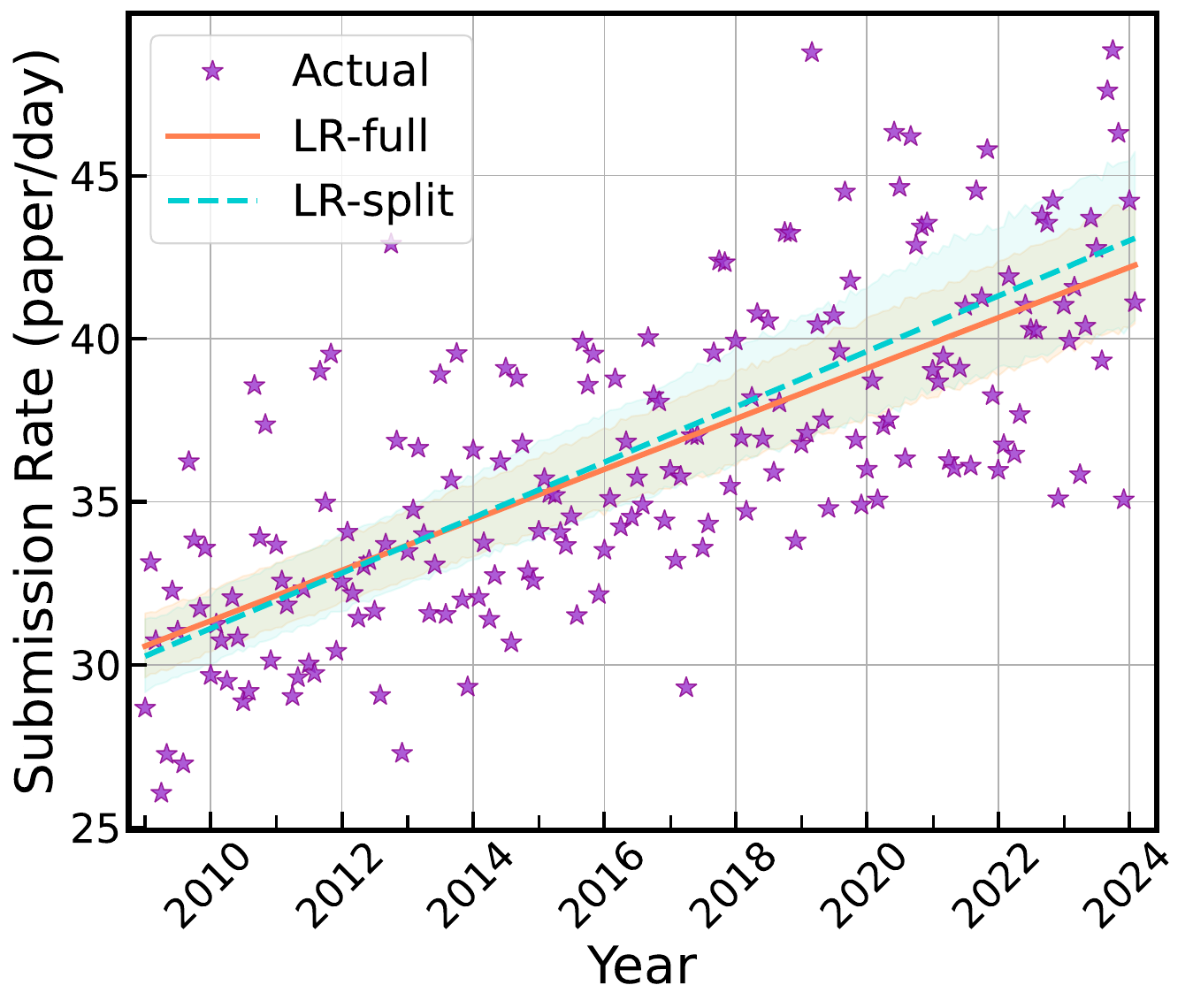}
\includegraphics[width=\columnwidth]{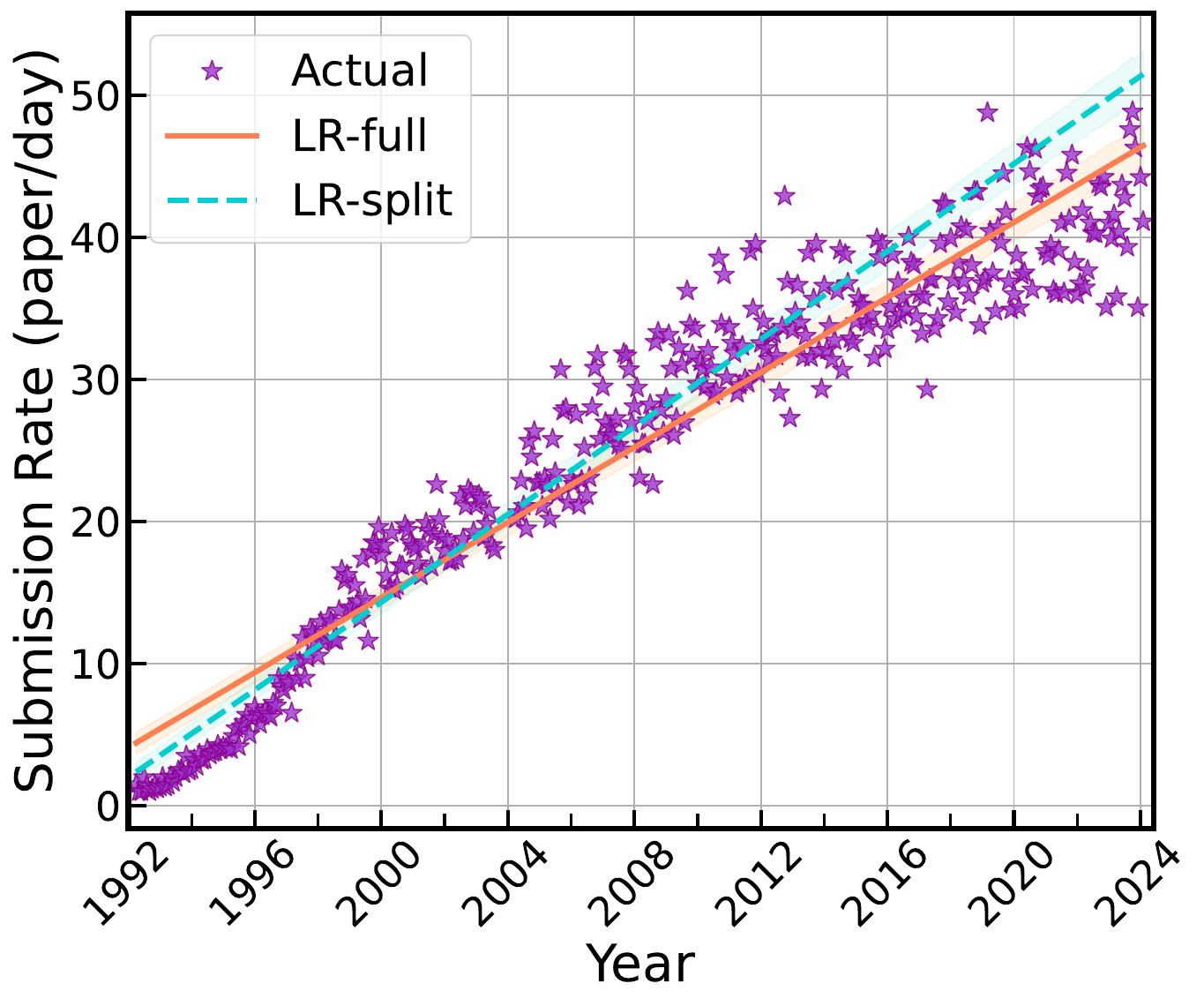}
\caption{\label{fig2:LR_fits_ALL_full} Same as Figure \ref{fig1:LR_fits_EP_HE}, but for \aph\ in aggregate since 2009+ (group 7, top panel) and for \aph\ as a whole since 1992+ (group 8, bottom panel).}
\end{figure}

Figure \ref{fig1:LR_fits_EP_HE} shows the LR-full and LR-split fits for the submission data of \aep\ (top panel) and of \ahe\ (bottom panel). Upon inspection, one notices that the slope of both LR-full fits is slightly smaller than that of the LR-split fits. I infer that the submission rate has increased slower in recent times. This is due to the fact that the LR-split is essentially a fit onto 4/5 of the dataset but applies the model onto the full dataset. Since the slope decreases when the LR model considers data from recent times, I deduce that the rates of expansion of these two \aph\ subcategories are decreasing. Unfortunately, the confidence intervals do not include most of the actual submission data. %, which is a sign of a few factors relevant to either the data or the model. 
I will discuss this observation in Section \ref{sec4:disc} and the possible explanations. Let's now shift our focus to Figure \ref{fig2:LR_fits_ALL_full}, which shows both the LR-full and LR-split fits for the aggregate submission data for \aph\ since 2009+ (top panel) and for \aph\ since 1992+. One notices a similar observation as Figure \ref{fig1:LR_fits_EP_HE}, in which the slope of the LR-full fits is smaller than that from the LR-split fits. The same deduction is made --- the rates of expansion of these two data groups are decreasing over time. Additionally, as I am using \aph\ as a whole to be a proxy of the \axvr, this leads to the inference that the rate of expansion of the \axvr\ is decreasing.

\section{The \axvr\ constant}
\label{sec3:a0 const}
The time is nigh to introduce the backbone of this work --- the \axvr\ constant, $a_0$. Essentially, it is defined as the rate of expansion of the \axvr. The slope from the LR fits delineated in Section \ref{subsec2-2:LR fits} describes the (monthly) rate of change in submission rate, and serves as the (almost\footnote{Including this in case someone comes at me and says that nothing is perfect, so here's to me covering all bases.}) perfect proxy for the rate of expansion of the \axvr, with units of paper/day/month. The slope from the LR fits for the \aph\ subcategories describes the rate of expansion of the respective sub-\axvr, while the slope from the LR fits for \aph\ as a whole would then be the proxy for the \axvr.

\begin{figure}[ht!]
\centering
\includegraphics[width=\linewidth]{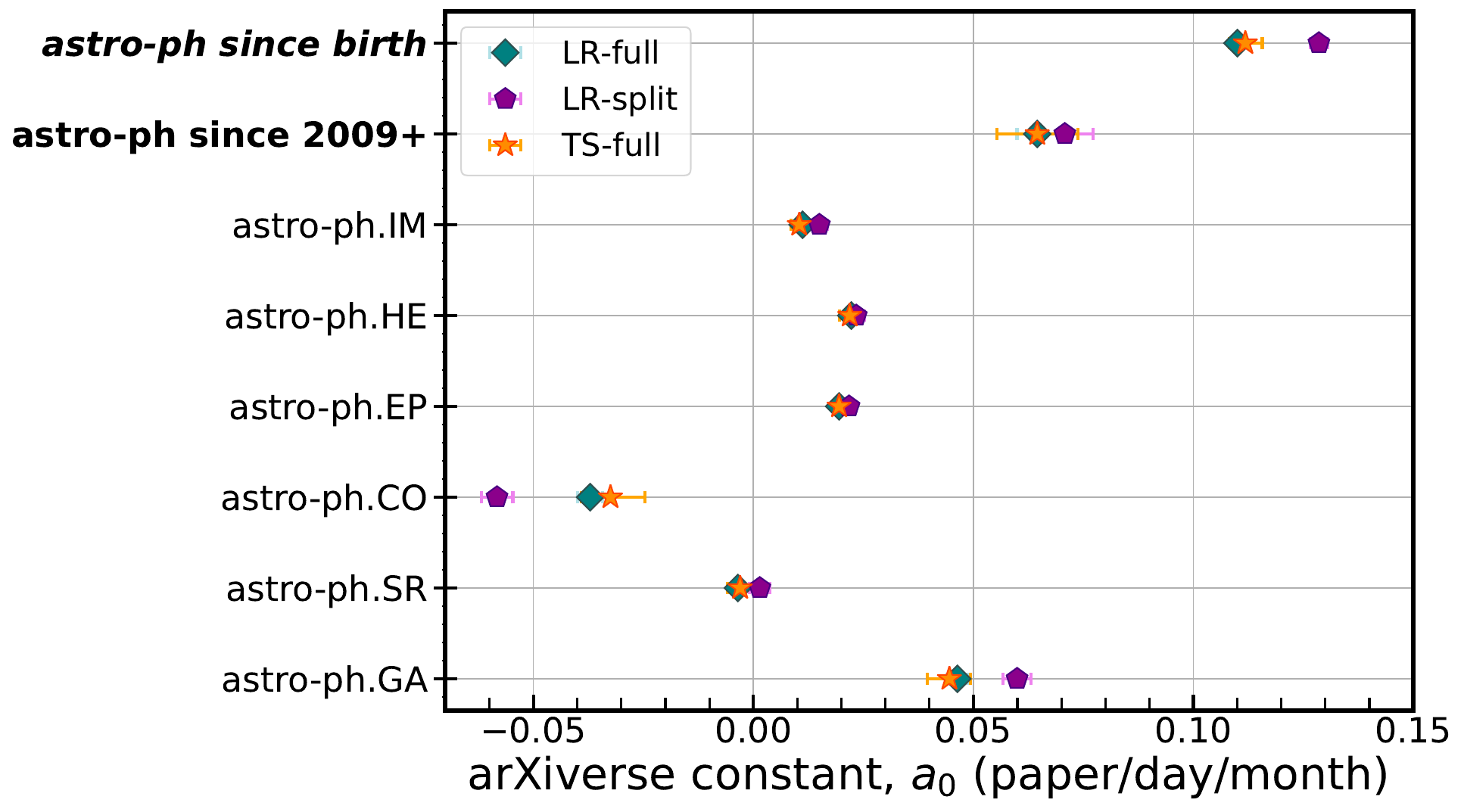}
\caption{\label{fig3:LR_slopes_comp} The \axvr\ constant, $a_0$, as measured using three different linear regression fittings on all data groups (labelled on y-axis). The teal diamonds indicate $a_0$ measured using the \texttt{scipy.stats.linregress} method on the full dataset, the purple pentagons indicate $a_0$ measured using also \texttt{scipy.stats.linregress} but on the split dataset, and the orange stars indicate $a_0$ measured using the Theil-Sen estimator on the full dataset.}
\end{figure}

Thus, for each data group, I have two $a_0$ measurements from the two LR fits (refer to Section \ref{subsec2-2:LR fits}). As I want to have another robust measurement of $a_0$ for each data group, I implement the Theil-Sen estimator (using \texttt{scipy.stats.theilslopes}) to compute the slope, the intercept, and the confidence interval of the slope of a robust LR fit to the complete dataset for a selected group. This method is referred to as TS-full hereafter. Figure \ref{fig3:LR_slopes_comp} compares the $a_0$ measured from the three LR fits. We can see that the $a_0$ measured from LR-full and TS-full agree very well for all eight groups. In contrast, LR-split only agrees with LR-full (or TS-full) for four \aph\ subcategories and also \aph\ in aggregate from 2009 onwards. The lack of consistent agreement is likely due to the split in datasets into training and testing sets. This also indicates that the measured $a_0$ changes over time. As observed in Figure \ref{fig2:LR_fits_ALL_full} where the slope measured and thus $a_0$ decreases over time, this is confirmed in Figure \ref{fig3:LR_slopes_comp}, where the $a_0$ for LR-full for both \aph\ in aggregate since 2009+ and for \aph\ in aggregate since birth (i.e., 1992+) is smaller than that for LR-split. While it may be very appealing to immediately jump to the conclusion that \axvr\ is then experiencing a deceleration in its rate of expansion, I thought it prudent to proceed with caution. Upon a quick glance, we can also observe that except for \aco\ and \asr, all data groups have positive $a_0$ measured. This indicates that while other \aph\ subcategories and even \aph\ as a whole may experience an increase in submission rate across time, this is not the case for \aco\ and \asr. These subcategories instead experience a decrease in submission rate and an almost constant submission rate, respectively. Therefore, we can deduce that not every sub-\axvr\ experiences the same rate of expansion, let alone the same change in rate of expansion. Thus, I conclude that there is a tension in the \axvr\ constant, which can vary depending on the dataset used for measurement. 

\begin{figure}
\centering
\includegraphics[width=\linewidth]{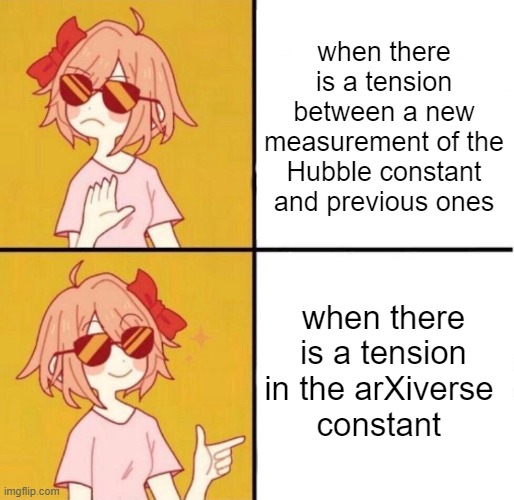}
\caption{\label{meem1:no yes} How we should feel about the \axvr\ tension.}
\end{figure}

At this stage, I think it is imperative to draw a clear line between the \axvr\ constant and a similar-sounding constant that we are probably very familiar with, i.e., the Hubble constant, $H_0$. While it may seem to readers that I am trying to draw a parallel comparison between the \axvr\ and the Universe, I am, in fact, not doing that. The \axvr\ constant is \textit{not} meant to be analogous to the Hubble constant as they are measures of distinct ``-verses" and are thus not comparable. Any coincidence is purely unintended \footnote{:>}.  
Anyway, here is a cartoon illustration for you. In Figure \ref{meem1:no yes}, I show how we should feel about the different tensions in cosmic measurements.

%\section{Results}

\section{Scrutinizing the results}
\label{sec4:disc}
\subsection{How does $a_0$ really change over time?}
\begin{figure*}[h!]
\centering
\includegraphics[width=\linewidth]{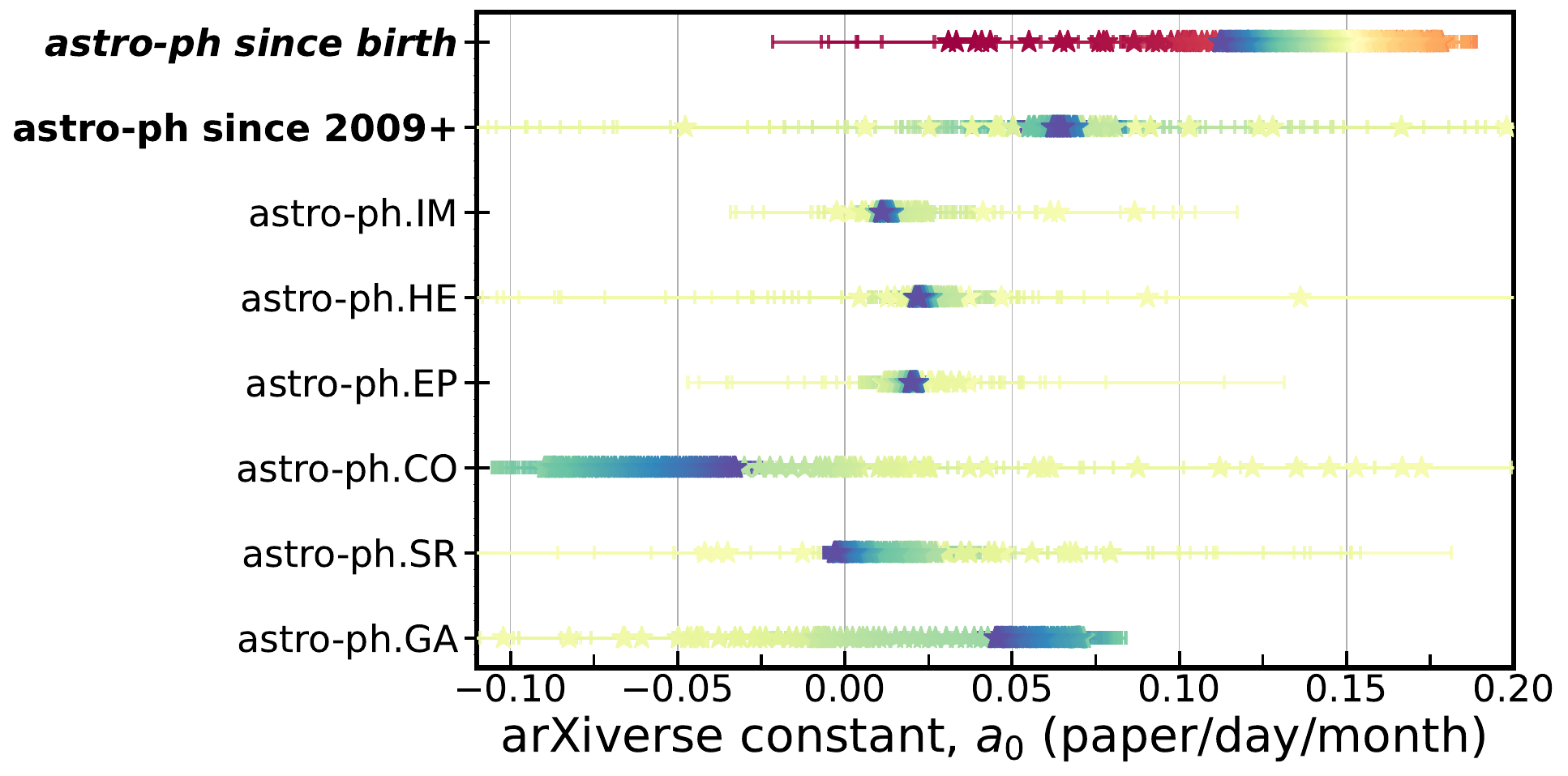}
\includegraphics[width=\linewidth]{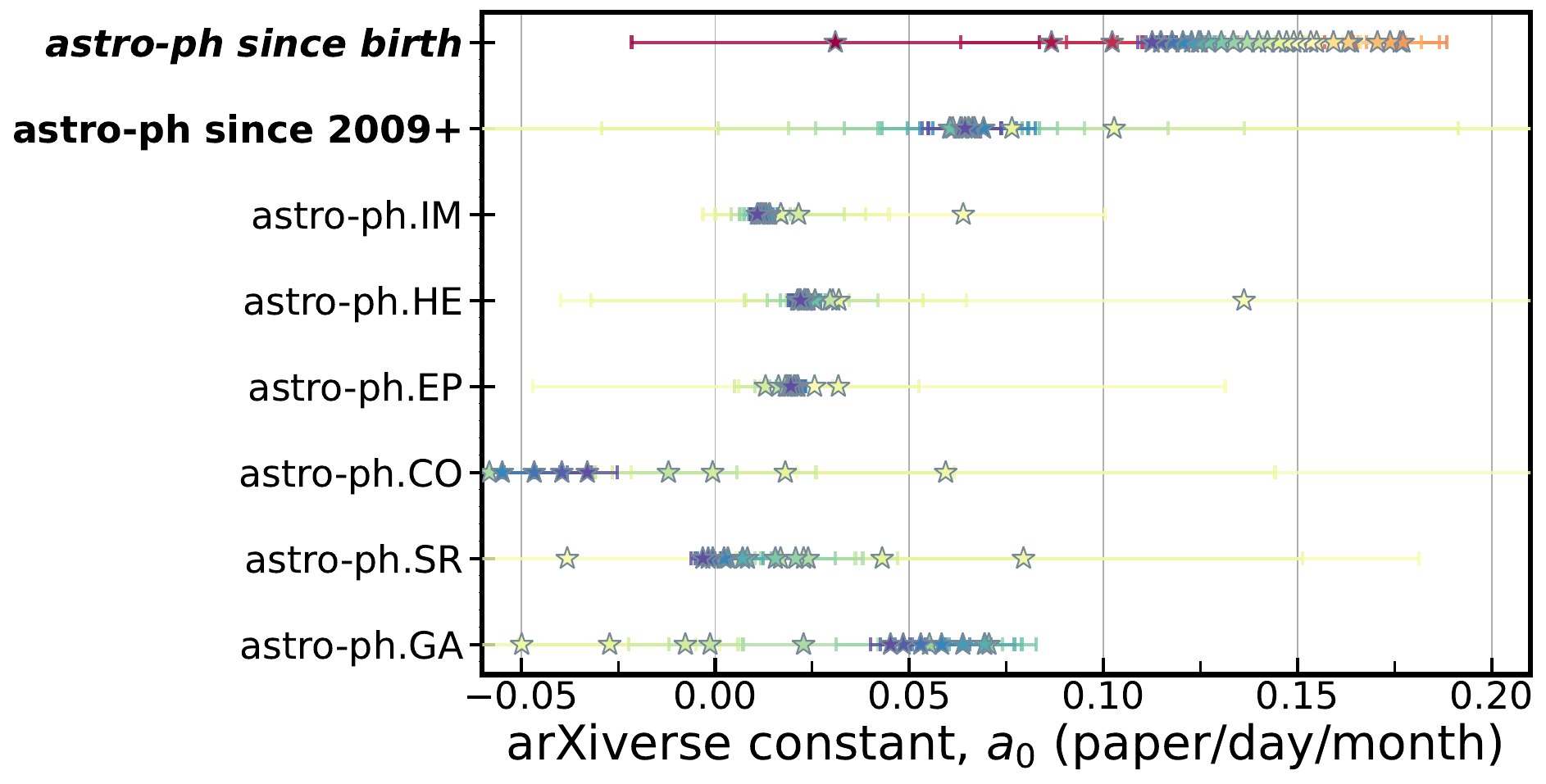}
\caption{\label{fig4:a0_comp} The \axvr\ constant, $a_0$, measured at a given point in time (from January 2010 for data groups 1--7, from April 1993 for group 8, see text for more information regarding dates, groups as defined in Section \ref{subsec2-1:yoink data}), using submission rate data up to selected point in time. Top panel shows $a_0$ measured every month while bottom panel shows $a_0$ measured every 12 months. The stars are colored using the \texttt{matplotlib} colormap `Spectral', going from red hues to blue hues akin to a rainbow's color spectrum, where the red end indicates earlier times (1992), blue ends indicate recent times (2024) and the yellow hues indicate the midpoint in time (\~{}2009!). }
\end{figure*}

In the preceding section, I establish that the $a_0$ changes over time and there is thus a tension in $a_0$ measurements. One way to verify this is to calculate the value of $a_0$ across time, since the birth of the \axvr\ (and the sub-\axvr \textsf{s}). For a given data group, I start with 12 data points, i.e., the first 12 months of submission data (average submitted paper per day), and implement the Theil-Sen estimator onto the selected data points to compute the slope and its confidence interval. The calculated slope and confidence interval then serve as the measured $a_0$ and its confidence interval at that point in time, which in this case, is the 13th month since I calculate the $a_0$ using all the available data up to that point in time. I then repeat the same process in the subsequent months, adding one data point to the Theil-Sen estimator at every iteration. This will then show us exactly how $a_0$ changes with time. I start with a minimum of 12 data points instead of fitting with fewer data points because I think having a year's worth of data is crucial for a robust initial measurement of $a_0$. For data groups 1--7, since the submission rate data starts from January 2009, this means that the first $a_0$ estimation was done in January 2010. For data group 8, the first estimation was done in April 1993. The $a_0$ measured across time using this method is shown in the top panel of Figure \ref{fig4:a0_comp}. I have split the \texttt{matplotlib} colormap `Spectral' into numerous hues to color-code the $a_0$ measurements across time. The colormap runs from red to blue hues akin to a rainbow's spectrum --- the red end of the colormap indicates earlier times (ca. 1992), while the blue end indicates later or recent times (ca. 2024). The midpoint of the spectrum is colored by yellow hues, which indicate half of \axv's lifetime to date. Coincidentally, the midpoint is around 2009, which was the point in time when \aph\ split into 6 subcategories and where we started having submission rate statistics for data groups 1--7. 

The top panel of Figure \ref{fig4:a0_comp} shows clearly that $a_0$ was rarely constant in time. Now, let's look into the values of $a_0$ over time for individual data groups. For \aga\ and \asr, there is an initial monotonic increase in $a_0$ measurements over time (stars with green hues having larger $a_0$ values compared to stars with yellow hues) and then a monotonic decrement in $a_0$ (indicated by stars with blue hues having smaller $a_0$ values than stars with green hues). In contrast, \aco\ shows a monotonic initial decrease, as indicated by the decrease in $a_0$ from the yellow-hued stars, to light-green stars, then to green stars. In more recent times, $a_0$ measured shows a monotonic increment, as indicated by the increase in $a_0$ from the green stars to the blue stars. These three subcategories solidify the notion that $a_0$ is not a constant in time. The three remaining subcategories, \aep, \ahe, and \aim, as well as \aph in aggregate since 2009+, initially have large fluctuations around the latest measurement of $a_0$, which gradually decreases over time, to small fluctuations around the final measurement. They show that while $a_0$ is not constant over time, it seems to settle in around a ``Goldilocks Zone" towards recent times. Lastly, when \aph\ is studied as a whole since its birth, it shows similar trends in $a_0$ as \aga\ and \asr\ -- there is an initial monotonic increment up to a certain point (stars with orange hues have larger values compared to stars with red hues) before $a_0$ starts to decrease monotonically ($a_0$ decreases from stars with orange hues to stars with blue hues).

I then simplify the $a_0$ measurement over time by measuring every subsequent 12 months instead of every subsequent month. This way, the potential seasonal effect in a year, which may cause fluctuations in $a_0$ measurements to a certain extent, can be absorbed into the LR fits. The $a_0$ measurements are shown in the bottom panel of Figure \ref{fig4:a0_comp}. The color codings are the same as the top panel. The trends in $a_0$ observed in the top panel for the eight data groups are also observed to be similar in the bottom panel, except that they are more evident in the bottom panel.

\begin{figure}[h!]
\centering
\includegraphics[width=\linewidth]{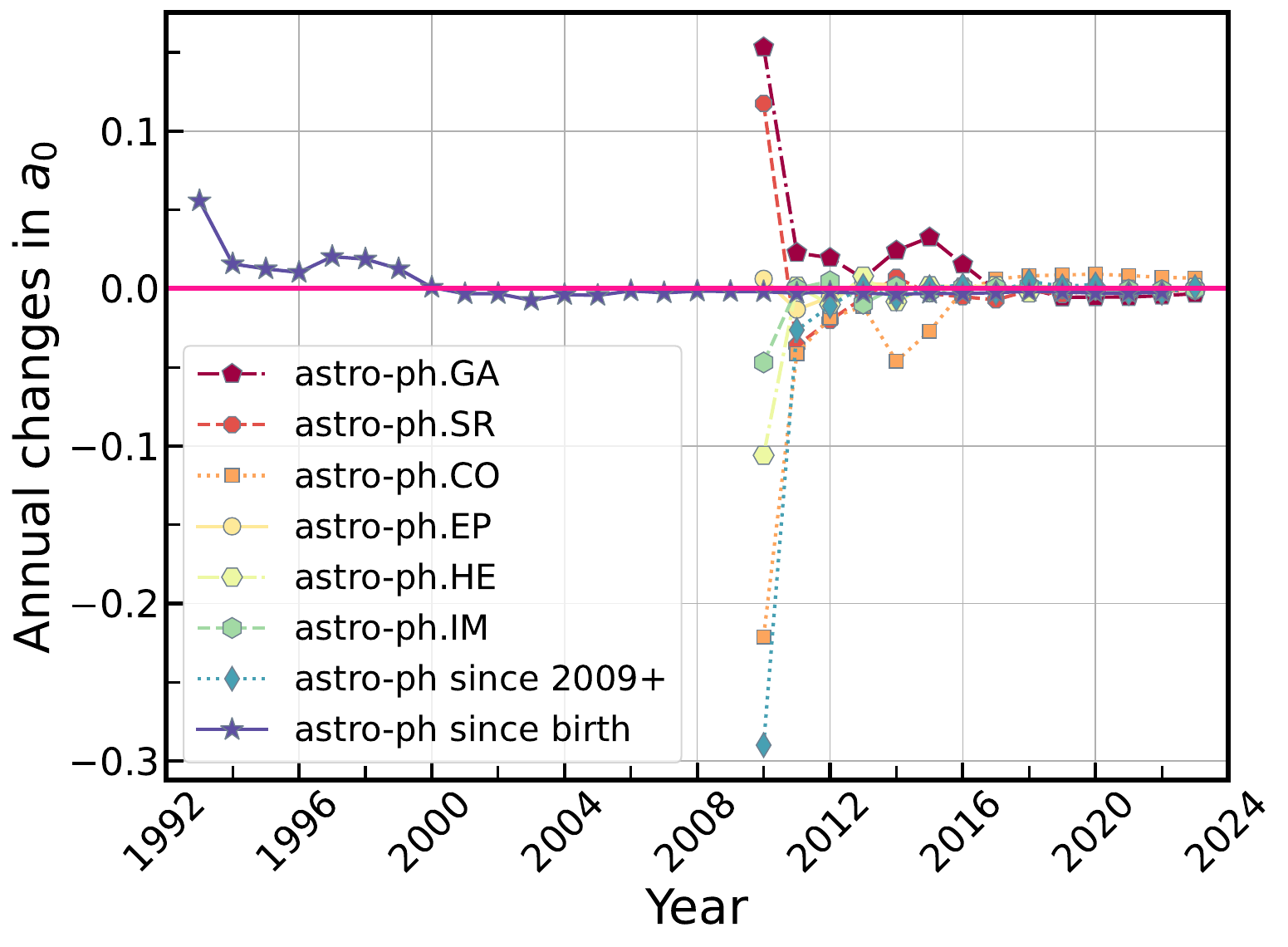}
\includegraphics[width=\linewidth]{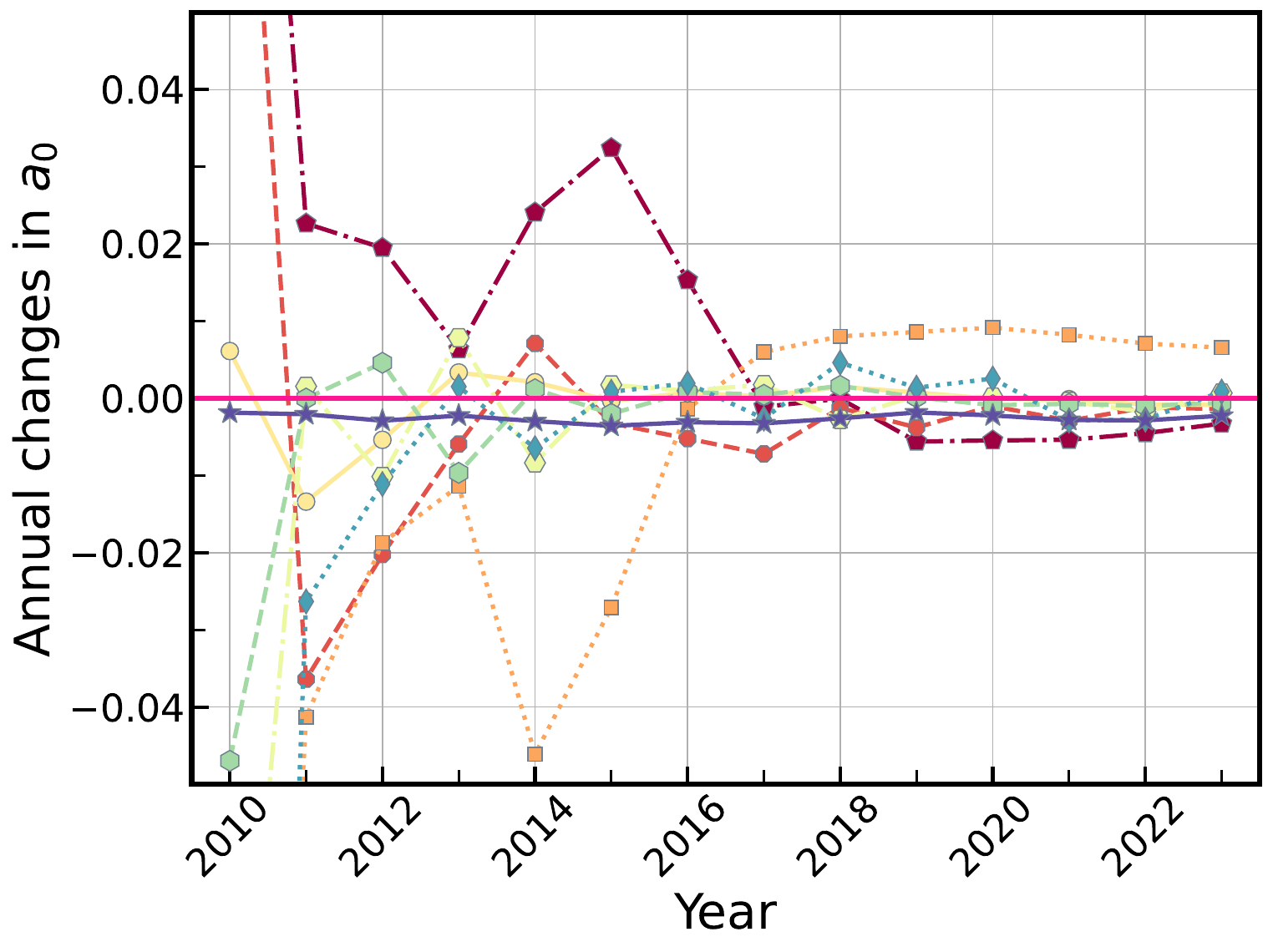}
\caption{\label{fig5:a0_diff} Annual changes in the \axvr\ constant, $\Delta\mathrm{a_0}$, for all 8 data groups. The top panel shows the full figure, running from 1992 to date. The bottom panel shows a zoomed-in version of the top panel, focusing on the changes in $a_0$ from 2009 onwards and at a smaller range of the y-axis. All 8 data groups show relatively larger fluctuations in $a_0$ at earlier times, which then tapers off in recent times. This suggests that the change in expansion rate of the \axvr\ is past its peak activity and is slowing.}
%From about 2017 onwards, the change in $a_0$ for all 8 data groups fluctuate minutely around 0, suggesting a slow change in the expansion rate of the \axvr.}
\end{figure}

\begin{figure}[ht!]
\centering
\includegraphics[width=\linewidth]{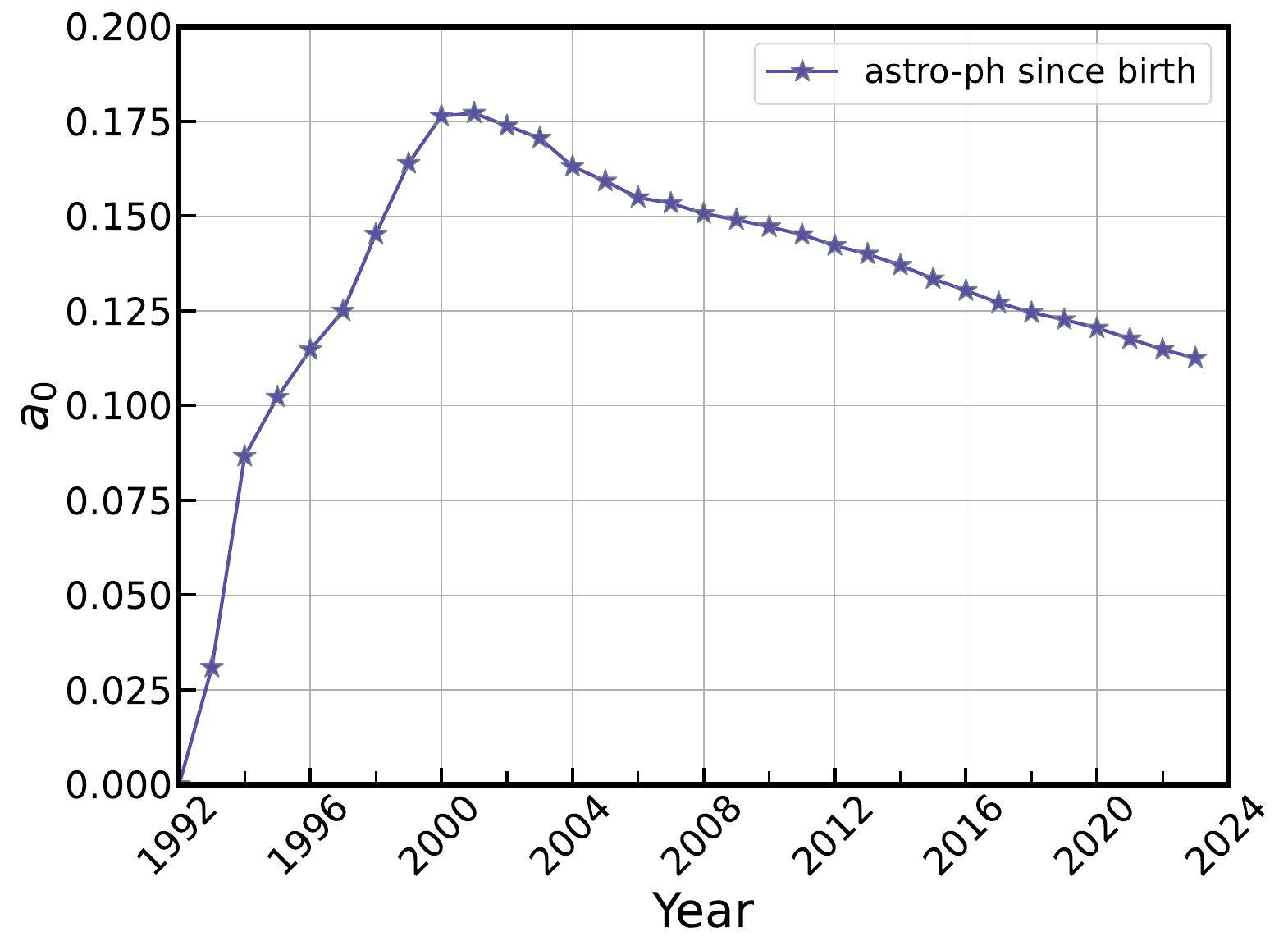}
\caption{\label{fig6:a0} The value of the \axvr\ constant, $a_0$, measured for the aggregate \aph\ since 1992+.}
\end{figure}

\begin{figure}[ht!]
\centering
\includegraphics[width=\linewidth]{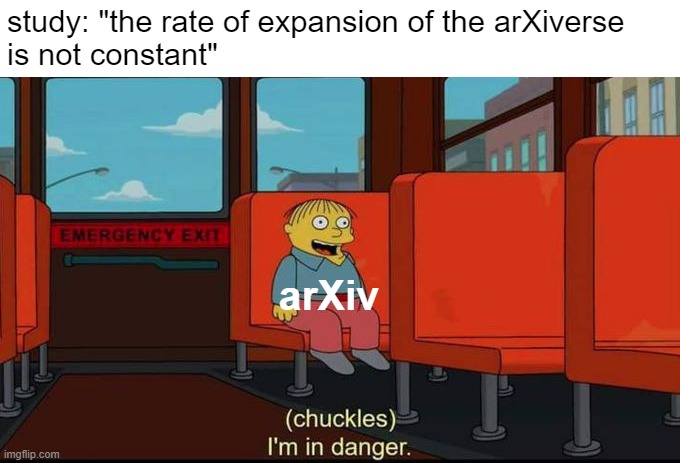}
\caption{\label{meem2:chuckle} How \axv\ presumably feels about their fate.}
\end{figure}

I then look into another metric of $a_0$ that could be useful, i.e., the annual change in $a_0$, or $\Delta\mathrm{a_0}$. This is essentially the year-to-year difference in the measured $a_0$ using the available data up to the selected point in time. The result is shown in Figure \ref{fig5:a0_diff}. The top panel of Figure \ref{fig5:a0_diff} shows $\Delta\mathrm{a_0}$ for the full temporal baseline -- from 1992 to date (from 2009 to date for data groups 1--7). I observe that all data groups, at the earlier times of their respective dataset, have relatively large fluctuations in $a_0$ compared to those of recent times. The bottom panel of Figure \ref{fig5:a0_diff} shows the zoomed-in version of the top panel, focusing on 2009 onwards at a smaller y-axis range. The hovering of $\Delta\mathrm{a_0}$ around zero from 2017 onwards for 
all the data groups (and from ca. 2000 onwards for group 8, as seen in the top panel) shows that $a_0$ is very likely to settle around a ``Goldilocks" value soon. It also suggests that about 8 years' worth of data is required for a more robust measurement of $a_0$, as that is the point of time where $\Delta\mathrm{a_0}$ starts being relatively smaller and tapers off. Upon closer inspection, the fluctuation of $\Delta\mathrm{a_0}$ around zero for \aep, \ahe, \aim, and \aph\ in aggregate since 2009 can be observed. The initial increment for both \aga\ and \asr\ before transitioning into a consistent decrease is also observed. For \aco, the initial decrease before switching over to a consistent increase is indicated as well. Lastly, for \aph\ as a whole since 1992, there is a consistent increment in $a_0$, marked by positive $\Delta\mathrm{a_0}$, until about 2000, before $\Delta\mathrm{a_0}$ is consistently below zero. In the grand scheme of things, we do want to study the \axvr\ as a whole since its birth instead of just parts of it, so at this point, \aph\ in aggregate since 1992+ is the best proxy for \axvr. Thus, the best proxy or measurement of the \axvr\ constant is the slope measured from \aph\ as a whole since 1992+, at the time of writing. Hence, $a_0$ is quantified at 0.110$\pm$0.002 paper/day/month. Last but not least, I deduce that the \axvr\ is past its peak of expansion (evident by the decelerating increase in $a_0$ in the earlier years) and is now experiencing a decreasing rate of expansion. This means that the \axvr, if nothing is done to alter the rate of expansion, could slow down to a crunch. I show a definitive\footnote{Questionable.} proof of this claim in Figure \ref{fig6:a0}. Let that sink in. In Figure \ref{meem2:chuckle}, I show an illustration of what I think \axv\ might be feeling upon hearing about their fate.

\subsection{Limitations and Future Prospects}
Throughout this work, I have performed LR fitting onto the datasets, which may be inadequate for the data. This is evident when one notices that the 95\% confidence intervals of the fits in both panels of Figures \ref{fig1:LR_fits_EP_HE} and \ref{fig2:LR_fits_ALL_full} do not cover most of the actual data points. This suggests that the LR models may be inadequate and underfit the data. It is possible that a polynomial or exponential regression may be more suitable for different data groups. Nevertheless, to ensure a significant correlation between the average submission rate and the passage of time, I compute the Pearson $p$-value for all eight data groups and found that except for \asr\ (which has a $p$-value of 0.026), all the other data groups have extremely low $p$-values. Since the significance level is typically set at 0.05, I reject the null hypothesis that there is no correlation between submission rate and time. In fact, the low $p$-values suggest that the observed correlation is statistically significant.

This work has used \aph\ as a whole as a proxy for \axvr. This choice is motivated by the preliminary nature of this study, the fact that \aph\ has been there almost since the birth of the \axvr, and that \aph\ has been one of the most popular categories on \axv. However, this is not the best decision since \axv\ has many more categories, even if \aph\ is one of the oldest or most popular categories. In the future, I propose to measure the $a_0$ using submission data from various \axv\ categories and their respective subcategories to obtain a more robust measurement of $a_0$, one that is more representative of the entire \axvr.

\section{Conclusion}
\label{sec6:conc}
In this work, I study the submission statistics of \axv\ \aph\ from 1992 to date. I fit linear regression models to the time series data of the monthly submission rate of \aph\ as a whole and for its six subcategories individually (\S\ref{sec2}). The slope of the LR fits, which tells us the monthly rate of change of submission rate, serves as the proxy for the rate of expansion of the \axvr. I coin the term ``the \axvr\ constant", $a_0$, to easier quantify the rate of expansion of the \axvr\ (\S\ref{sec3:a0 const}). From the measured slopes as proxies for $a_0$, I find that \aph\ as a whole (both since birth or since its split into six subcategories), \aga, \aep, \ahe, and \aim\ have positive $a_0$ --- their respective submission rate (or rate of expansion) increases over time. On the other hand, \aco\ has negative $a_0$ while \asr\ has near-zero $a_0$. This suggests that their submission rates decrease over time and stay constant over time, respectively. The very fact that $a_0$ is different across subcategories (and data groups) suggests that the rates of expansion of the (sub-)\axvr\textsf{s} are different from each other. This indicates that there is a tension in the \axvr\ constant, and that $a_0$ changes over time.

To further support the inference that $a_0$ is not really a constant in time, I measure $a_0$ across time for all the data groups (\S\ref{sec4:disc}). The measured $a_0$ over time shows that it indeed changes at every given point of time. Four of the data groups show indications that they might be settling into a ``Goldilocks Zone" of $a_0$ since the measured $a_0$ have been hovering right around the final measurement of $a_0$. Three of the data groups indicate that there is an initial increment in $a_0$ before a consistent decrease in recent times; one data group shows the opposite --- it has an initial decrement in $a_0$ before slowly increasing in recent times. I then calculate the year-to-year difference in $a_0$, $\Delta\mathrm{a_0}$, and find the same trends in $a_0$ as described earlier in this paragraph. Nevertheless, there is more work to be done in the same spirit. I aim to implement various regression models onto the submission statistics and expand the dataset used for the regression modeling. I believe these two implementations will increase the robustness of the measured $a_0$.

\begin{figure}[ht!]
\centering
\includegraphics[width=\linewidth]{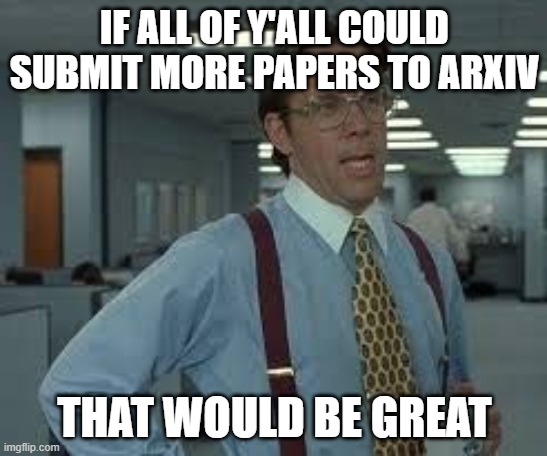}
\caption{\label{meem3:pls} Call to action.}
\end{figure}

In conclusion, there is an evident tension in the \axvr\ constant measured across various datasets. This is marked by the different measurements of $a_0$ depending on the dataset and temporal baseline used. Having said that, the best proxy for the \axvr\ constant that I have at the moment is the $a_0$ measured from \aph\ as a whole since 1992+. This is because it measures the rate of expansion of \aph\ since its birth (and also the birth of \axv). Thus, in March 2024, $a_0$ is measured to be at 0.110$\pm$0.002 paper/day/month. With that said, $a_0$ has been consistently decreasing over time; this decrease in rate of expansion indicates that the \axvr\ could slow down to a crunch. It is vital for us to play our part in ensuring that the \axvr\ does not suffer this painful fate. Collectively, we can reverse this by yeeting more papers to \axv. We all hold the power in our pens and papers to decide \axvr's fate. The fate is written in your hands (and perhaps the stars too!). Figure \ref{meem3:pls} shows a call to action for this purpose. Remember, the pen (and paper) are mightier than the sword.

\section*{Acknowledgements} 
\noindent The author thanks Irina Ene for the most wonderful and memeingful discussions that made this research so enjoyable and exciting. The author also thanks peer reviewers, Hitesh Kishore Das and Anshuman Acharya, for their kind efforts in reviewing this paper and ensuring the science here is logical and sound. They will be beer-ly rewarded for their contributions. \textit{``Do a peer review, get a beer reward,"} a wise person once said. This research has made use of arXiv API, and the author would like to thank arXiv for the use of its open-access interoperability.
%The authors acknowledge the time and effort researchers invested in executing original research, presenting methods and findings in papers, as well as embellishing the papers with eye-catching titles. They are not the easiest of tasks. We also thank the entertaining and/or eyeroll-inducing moments just from reading the paper titles -- it was a nice change of pace in our mundane life. We thank you for your creativity :) %We also acknowledge Matplotlib \citep{Hunter:2007}, Numpy \citep{harris2020array}, and Scipy \citep{2020SciPy-NMeth}.
This research has also made use of 
\software{Matplotlib \citep{matplotlib}, SciPy \citep{2020SciPy-NMeth}, NumPy \citep{numpy} and scikit-learn \citep{scikit-learn}}.

%\begin{acknowledgements}
%\end{acknowledgements}

\bibliography{biblio}{}
\bibliographystyle{aasjournal}

\end{CJK*}
\end{document}